# When Automatic Voice Disguise Meets Automatic Speaker Verification

Linlin Zheng, Jiakang Li, Meng Sun, Xiongwei Zhang, and Thomas Fang Zheng, *Senior Member, IEEE*

*Abstract*—The technique of transforming voices in order to hide the real identity of a speaker is called voice disguise, among which automatic voice disguise (AVD) by modifying the spectral and temporal characteristics of voices with miscellaneous algorithms are easily conducted with softwares accessible to the public. AVD has posed great threat to both human listening and automatic speaker verification (ASV). In this paper, we have found that ASV is not only a victim of AVD but could be a tool to beat some simple types of AVD. Firstly, three types of AVD, pitch scaling, vocal tract length normalization (VTLN) and voice conversion (VC), are introduced as representative methods. State-of-the-art ASV methods are subsequently utilized to objectively evaluate the impact of AVD on ASV by equal error rates (EER). Moreover, an approach to restore disguised voice to its original version is proposed by minimizing a function of ASV scores w.r.t. restoration parameters. Experiments are then conducted on disguised voices from Voxceleb, a dataset recorded in real-world noisy scenario. The results have shown that, for the voice disguise by pitch scaling, the proposed approach obtains an EER around 7% comparing to the 30% EER of a recently proposed baseline using the ratio of fundamental frequencies. The proposed approach generalizes well to restore the disguise with nonlinear frequency warping in VTLN by reducing its EER from 34.3% to 18.5%. However, it is difficult to restore the source speakers in VC by our approach, where more complex forms of restoration functions or other paralinguistic cues might be necessary to restore the nonlinear transform in VC. Finally, contrastive visualization on ASV features with and without restoration illustrate the role of the proposed approach in an intuitive way.

## I. INTRODUCTION

In the era of artificial intelligence and big data, Automatic speaker verification (ASV) is an important means to automate the surveillance of fraud calls. ASV makes a decision based on whether two pieces of voices are from the same person or not, by extracting their features and computing their similarities. However, the attackers could hide their real identities by voice disguise. Consequently, ASV would fail when making decisions with the disguised voices.

Automatic voice disguise (AVD) aims to automatically transform voices by algorithms in order to hide the real identity of a speaker, and can be easily conducted with softwares accessible to the public and has brought great threat to information security. For example, pitch scaling can raise or decrease the pitch of a piece of voice by frequency shifting or temporal stretching in a linear style [1]. Improving from the linear transform to nonlinear ones, Vocal Tract Length Normalization (VTLN) warps the frequency axis into different directions [2]. Theoretically, any function from $[0, \pi]$ to $[0, \pi]$ can perform as the frequency warping in VTLN, if the warping retains the intelligibility of the transformed speech. Voice conversion (VC) refers to digital cloning of a person's voice, which can be used to modify audio waveform so that it appears as if spoken by some specific person (i.e. the target speaker) than the original speaker (i.e. the source speaker). Recently, voice conversion challenges have been conducted to elaborate VC techniques in VCC 2016, 2018 and 2020 [3]. With the nonlinear mapping of spectral features in VC, the identity of the source speaker could be hidden since the converted voice sounds like from the target speaker. In summary, all these AVD methods would mislead ASV by modifying the spectral and temporal characteristics of voices, as will be quantitatively evaluated in Section VI.A.

With the increasing of crimes conducted by voice disguise, a lot of research has explored the impacts of disguised voices on ASV. The research in [4]-[6] revealed the vulnerability of traditional ASV systems against human disguise of voices. The work in [7] reported the Equal Error Rate (EER) on human disguised voice was as high as 24.7% by using a recently proposed deep neural network method for ASV from [8]. The work in [4] revealed that the performance of human disguise depended on the disguising skills of the imposter. Therefore, non-skilled imposters may tend to use AVD instead, rather than human disguise. It is thus important to evaluate the impact of AVD on state-of-the-art ASV (e.g. *x*-vectors [9]). Recently, [10] reported the severe deterioration of the performance of *x*-vectors on VC-based speaker anonymization. All these works have revealed the vulnerability of ASV on voice disguise.

The first step to defend against AVD is detecting whether a piece of voice is disguised or not [11]. Algorithms by analyzing the changing rule of the Mel Frequency Cepstral Coefficients

L. Zheng, J. Li, M. Sun, X. Zhang are with the Laboratory of Intelligent Information Processing, Army Engineering University, Nanjing, China. T. F. Zheng is with the Center for Speech and Language Technologies (CSLT), Tsinghua University, Beijing, China.
L. Zheng and J. Li contributed equally to this work.
Corresponding: M. Sun, e-mail: sunmengccjs@gmail.com.
This work was supported by the Natural Science Foundation of Jiangsu Province under Grant BK20180080 and the National Natural Science Foundation of China under Grant 61471394.

(MFCC) through Gaussian Mixture Models (GMM) were investigated to identify disguised voices by pitch scaling in [12]. MFCCs and their statistical moments were extracted and utilized as input features for Support Vector Machines (SVM) to classify voices into normal ones or disguised ones by pitch scaling [13]-[15]. The experiments showed that the detection rate reached over 90% accuracy on their cross-database evaluation. A method based on a dense convolutional network for detecting disguised voice by pitch scaling from genuine voice was presented in [16]. The experimental results showed that the average accuracies over intra-database and cross-database is 96.45%, superior to the state-of-the-art methods. Given the fact that automatically disguised voices should be synthesized and replayed to attack ASV systems, the detection of speech synthesis and voice conversion (named by logical attack, LA), and the detection of recording replay (named by physical attack, PA), were proposed and extensively studied in ASVspoof 2015, 2017 and 2019 [17]-[19]. The proposed anti-spoofing approach had successfully detected spoofing voices from speech synthesis, voice conversion, and recording replay etc., with high accuracies. These techniques usually perform as a prerequisite step to secure ASV systems by filtering out spoofing voices, which hence exclude disguised ones. Data augmentation was utilized in [10] to link disguised voices and their original ones by re-training *i*-vectors/*x*-vectors in the training and enrollment stages of ASV for each speaker. In this paper, the AVD detection and speaker verification will be merged, where normal voice will be treated as a special case of disguised voice with some special disguising parameters. Therefore, the detection of AVD does not have to be a standalone task any more, which is an advantage of the proposed method in this paper.

It seems not enough for tracing the criminals conducting voice disguise by only classifying a piece of voice is disguised or not, but the restoration of the original voice is necessary for listening tests as interpretable evidences. In order to improve ASV on disguised voices, Dynamic Time Warping (DTW) was applied to restore pitch scaling disguised voices by estimating the degree of disguise, which improved the performance of Vector Quantization (VQ) on speaker recognition [20]. The ratio of fundamental frequencies was utilized to estimate the degree of pitch-scaling disguise for voice restoration, based on which restored MFCCs were extracted as features of a GMM-UBM system for ASV in [21]. In their objective evaluation, the method yielded an EER around 4% on TIMIT, a dataset with clean speech, which significantly reduced the EER of 40% without the restoration of disguised voices. The methods proposed in [20] and [21] proved the necessity and usefulness of the restoration of disguised voices. However, the methods were only experimented on clean speech with traditional speaker recognition methods for pitch scaling. More challenging datasets, more complex disguising methods and more advanced speaker recognition models should be considered to study the restoration of disguised voices, which is one of the motivating aspect for this paper.

Furthermore, it is interesting to explore the mutual relation of ASV and AVD, e.g., why ASV is vulnerable to AVD, if the features and metrics involved in ASV reflect some hidden facts of the disguised voices, and if there exists a universal method to reduce the vulnerability of ASV on AVD. This is the other motivating aspect of this paper.

In this paper, AVD is firstly summarized into a function described by disguising parameters in Section II. State-of-the-art ASV methods are then introduced to quantitively evaluate the impact of AVD in Section III. An approach to estimate the restoration function is proposed by minimizing a function of ASV scores w.r.t. restoration parameters in Section IV. Experimental setups of datasets, baselines and the details of model configurations are presented in Section V. Results, comparison and discussion are given in Section VI. Finally, we conclude that ASV is not only a victim of voice disguise but could be a tool to beat AVD with relatively simple transformations in Section VII.

## II. MODELING OF AUTOMATIC VOICE DISGUISE

AVD is normally realized by modifying the temporal or spectral properties of voices, e.g. in a linear style by pitch scaling, in a non-linear style by VTLN or by a complex spectral mapping function in VC. In this section, we will briefly introduce these three types of disguise and analyze their pros and cons.

### A. Pitch scaling by frequency shifting or temporal stretching

Pitch scaling aims to modify the voice pitch of a speaker to hide his/her identity. In essence, it can be achieved by stretching the spectrum in frequency-domain or voice resampling in time-domain.

*1) Frequency-domain disguise*

Frequency-domain disguise is usually operated by expanding or compressing the spectrum while keeping the content of the voice unchanged. Mathematically, the instantaneous frequency $\omega$ is modified to $\omega'$ by introducing a scaling factor $s$, as defined in (1),

$$\omega' = s\omega, 0 \leq \omega', \omega \leq \frac{\pi}{2}. \quad (1)$$

Correspondingly, the disguised spectrum is,
$$F'(\omega') = F'(s\omega) = F(\omega). \quad (2)$$

It can be seen from the analysis above that the scaling factor $s$ plays a deterministic role in the frequency-domain disguise by stretching the spectrum.

*2) Time-domain disguise*

Time-domain disguise can be realized by adjusting the sampling rate, which changes the fundamental frequency of speech signal and hence the pitch [22]. However, the disguised voice generated in this way often sounds unnatural. A technique from speech synthesis, Pitch-Synchronous Overlap and Add method (PSOLA) is deployed to improve the naturalness the disguised voices [23]. Unvoiced speech is non-periodic and is quite close to white noise, which has little information on speaker identity and does not need to be transformed for voice disguising purpose. Assuming $x(t)$ is voiced speech, PSOLA first detects the position and contour of the formants of the signal by extracting the pitch-period parts. The time indices of a pitch-period are then modified by,

$$t' = \frac{t}{s}, 0 \leq t \leq P(t), 0 \leq t' \leq P'(t), \quad (3)$$

where $s$ is a scaling factor, $P'(t)$ is the modified pitch period $P(t)/s$. Therefore, the disguised waveform is obtained by,

$$x'(t') = x'\left(\frac{t}{s}\right) = x(t). \quad (4)$$

As same as in frequency-domain disguise, it can be seen from the analysis above that the scaling factor $s$ also plays a deterministic role in the time-domain disguise by stretching pitch-periods. PSOLA only modifies the prosodic features of fundamental frequency, duration, and short-term energy of speech, so that the disguised voice has a similar envelope as the original one.

*3) Unified representation for the disguise with pitch scaling*

In the technique of audio processing, pitch can be increased or decreased by up to 12 semitones [14]. Given this rule, the pitch of the original voice, $p_0$, and the pitch of the disguised voice, $p_1$, can be expressed by,

$$p_1 = 2^{\alpha/12} p_0, \quad (5)$$

where $\alpha$ is the semitone factor which is also called disguising parameter in this paper. It is straightforward to see that a positive/negative $\alpha$ turns up/down the pitch by $\alpha$ semitones, while $\alpha=0$ means no change on the voice.

By considering pitch as a special value of $\omega$ in (1), it is easy to find that,

$$s = \frac{\omega'}{\omega} \approx \frac{p_1}{p_0}, \quad (6)$$

which bridges the scaling factor $s$ and the pitch values of the original and disguised voices. The disguising parameter $\alpha$ and the scaling factor $s$ have thus the following relationship,

$$\alpha \approx 12 \log_2(s). \quad (7)$$

As explained above, AVD based on both frequency-domain disguise and time-domain disguise boils down to the proportional modification of frequency indices and pitch-periods, where scaling factor $s$ is introduced to quantize the transform. With the transform from $s$ to $\alpha$ in (7), the disguised voice $y$ is hereby represented in a unified form,

$$y = f(x; \alpha), \quad (8)$$

where $\alpha$ quantifies the frequency warping in (1) or the pitch-period stretching in (3) measured by semitones.

*B. Frequency warping by nonlinear functions in VTLN*

It is believed that the difference of the speaker's vocal tract length leads to the variability of the speech waveform of the same language content [2]. The original purpose of VTLN is to normalize the speaker's voice in order to remove individual speaker characteristics and improve the accuracy of speech recognition. VTLN can also be used for voice disguise by adjusting the frequency axis of the spectrogram through the warping function to hide the individual characteristics of the vocal tract length.

In this paper, AVD by VTLN has six steps: pitch marking, frame segmentation, fast Fourier transform (FFT), VTLN, inverse FFT and PSOLA [24]. The purpose of pitch marking and frame segmentation is to cut speech signals into frames that match the pseudo-periodic of a voiced sound, so that the output speech would have the best sound quality. In AVD by VTLN, the choice of frequency warping function plays an important role, where the commonly used functions have the following ones, see mathematical formulas in [2] and codes in [25].

*1) Bilinear function*

$$z' = \frac{z - \alpha}{1 - \alpha z} \text{ with } z = e^{i\omega}, z' = e^{i\omega'}, \quad (9)$$

*2) Quadratic function*

$$\omega' = \omega + \alpha \left(\frac{\omega}{\pi} - \left(\frac{\omega}{\pi}\right)^2\right), \quad (10)$$

*3) Power function*

$$\omega' = \left(\frac{\omega}{\pi}\right)^\alpha, \quad (11)$$

*4) Piecewise linear function*

$$\omega' = \begin{cases} (\alpha+1)\omega & \omega \leq \omega_0 \\ (\alpha+1)\omega_0 + \frac{\pi - (\alpha+1)\omega_0}{\pi - \omega_0}(\omega - \omega_0) & \omega \geq \omega_0 \end{cases},$$

$$\omega_0 = \begin{cases} \frac{7}{8}\pi & \alpha \leq 1 \\ \frac{7}{8(\alpha+1)}\pi & \alpha > 1 \end{cases}, \quad (12)$$

where $\omega$ is the original instantaneous frequency, $\omega'$ is the warped one, $\alpha$ is the disguising parameter to reflect the nonlinearity of the warping, and $i$ is the imaginary unit in (9).

For simplicity, the disguised voice $y$ by using the four warping functions of VTLN can also be summarized by (8) where different warping functions result in different $f$'s in (8) and $\alpha$ controls the disguising strength.

*C. Voice conversion by spectral mapping*

VC modifies a person's voice to imitate some target person while keeping the linguistic information unchanged. It is worth noting that voice conversion is different from voice transformation (e.g. pitch scaling in Section II.A and VTLN in Section II.B) where the former one has a specific target speaker while the latter one does not.

In this paper, the competitive baseline provided in VCC2018 is adopted and introduced here. A vocoder, e.g. STRAIGHT [26] or WORLD [27], is conventionally first utilized to decompose speech into F0 contour, spectral features, and aperiodic residuals, where the F0 contour and spectral features are converted to those of the target speakers while the aperiodic residuals does not have to be converted. The F0 contour is converted in its log linear format to match the statistics of the target speaker [3]. In the training stage, VC learns a mapping of spectral features between the source speaker and the target speaker. With the learned mappings, any voice from the source speaker can be converted to that sounds like from the target speaker.

Gaussian mixture models (GMM), deep neural network (DNN) and its variants, can be used to model the mapping of spectral features. In this paper, a strong VCC2018 baseline, differential GMM, is used for VC to perform spectral mapping, given its accessibility and reproducibility [28].

To be consistent with the representation of pitch scaling and VTLN, VC disguise can also be expressed by (8), where $\alpha$ is no longer a parameter but should be a group of parameters of GMM or DNN specified by the target speaker.

*D. Comments on the disguising methods*

In practice, VC generally requires a large amount of data from many target persons to train different VC models to make sufficient confusion on the disguised voices (e.g. by switching the target person randomly). This brings certain difficulties and costs in implementation. In contrast, voice transform (such as pitch scaling and VTLN) does not require any additional data from target speakers and is easier to implement than VC, which makes pitch scaling and VTLN being easily integrated to many popular audio editing softwares [16].

In summary, when choosing disguising methods, there is usually a trade-off among the performance on deceiving ASV systems, the additional computational cost, and the robustness on voice quality. Given the comparison on the three disguising methods, pitch scaling seems a good compromise on the three points above, as will be presented in Section VI.F.

## III. EVALUATION OF AUTOMATIC VOICE DISGUISE BY STATE-OF-THE-ART ASV

As a conventional ASV method, GMM-UBM had been utilized to evaluate the impact of AVD in [21]. In the past decades, ASV has reached great progress from *i*-vectors [29] to *x*-vectors [9] and many other deep learning models. In this section, state-of-the-art ASV with Time-Delay Neural Networks (TDNN) [30] and Additive Margin Softmax (AMSoftmax) [31] model is briefly introduced, whose usage on evaluating voice disguise will be presented.

*A. ASV with TDNN and AMSoftmax*

*x*-vectors extracted from TDNN have significantly improved the accuracy of text-independent speaker verification [9][32]. TDNN with AMSoftmax further improves the performance of *x*-vectors by maximizing the between-class margin and minimizing the within-class margin [33]-[35]. The conceptual graph of training TDNN with AMSoftmax loss is shown in Fig.1, where TDNN transforms an utterance into a vector, while AMSoftmax aims at enlarging the distance of vectors from *anchor* and *negative* utterances and shrinking the distance of vectors from *anchor* and *positive* utterances.

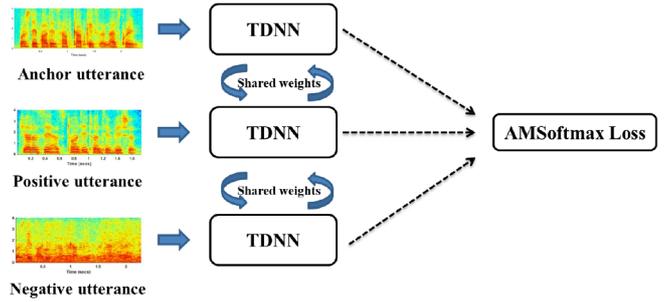

FIG. 1: TRAINING OF TDNN WITH AMSOFTMAX LOSS. ANCHOR AND POSITIVE UTTERANCES ARE FROM THE SAME SPEAKER, WHILE ANCHOR AND NEGATIVE UTTERANCES ARE FROM DIFFERENT SPEAKERS. THE ROLE OF AMSOFTMAX LOSS IS TO PULL THE 'ANCHOR-POSITIVE' PAIR CLOSER AND TO PUSH THE 'ANCHOR-NEGATIVE' FURTHER.

The TDNN architecture of the ASV in this paper is the same as that in [34] and is shown in Fig.2. Speaker embedding vectors are firstly extracted by layer-6 (i.e. $l_6$ in Fig.2) of TDNN. *x*-vectors are subsequently obtained by making dimension reduction on the speaker embedding vectors using Linear Discriminative Analysis (LDA). By denoting the network layers from input to $l_6$ and the LDA as a function $g(\cdot)$, *x*-vector can be represented by $g(x)$ and $g(y)$ for original voice $x$ and disguised voice $y$, respectively. The verification task thus boils down to computing the distance, $d(g(x),g(y))$, where $d$ is a metric, e.g. probabilistic LDA (PLDA), or cosine distance, which may be computed by some standard recipe in ASV [36].

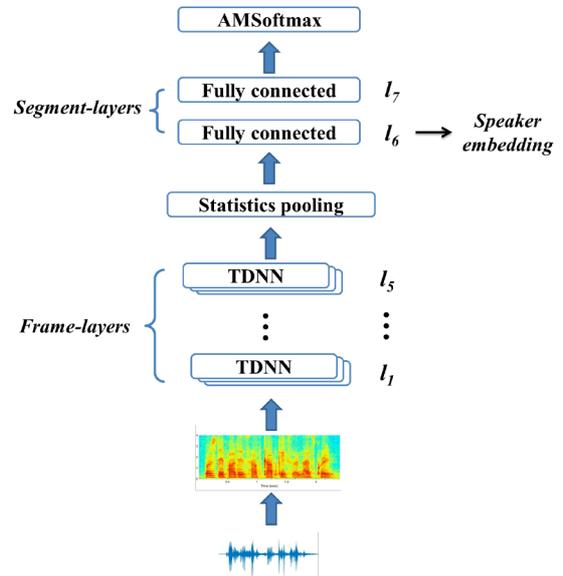

FIG. 2: ILLUSTRATION OF TDNN AND SPEAKER EMBEDDING. THE NETWORK HAS SEVEN LAYERS IN TOTAL WITH FIVE FRAME-LAYERS AND TWO SEGMENT-LAYERS. STATISTICAL POOLING IS APPLIED TO TRANSFORM UTTERANCES WITH ARBITRARY LENGTH TO A UNIFORM LENGTH. THE SEGMENT-LAYER 6 IS KNOWN AS SPEAKER EMBEDDING.

*B. Evaluation of AVD by EERs quantitatively*

A list of trials is considered for evaluation, where each trial consists of a pair $\{x_i, y_{j,\alpha}\}$. $x_i$ is a piece of normal voice from speaker $i$ ($1 \leq i \leq C$), $y_{j,\alpha}$ is a piece of disguised voice from speaker

$j$ ($1 \leq j \leq C$) with disguising parameter $\alpha$ (for pitch scaling and VTLN, $\alpha$ is sampled from an interval $-R \leq \alpha \leq R$; while for VC, $\alpha$ is a pre-trained GMM or DNN model for a target speaker), and $C$ is the total number of speakers for testing. The information on $i$, $j$ and $\alpha$ is only used for making testing datasets, on which the ASV algorithms are completely blind.

The values of $i$ and $j$ are both random samples to make a representative list of trials. Speaker $i$ can be either the same as speaker $j$ or not. A decision on if $x_i$ and $y_{j,\alpha}$ are from the same speaker is made by introducing a threshold $\eta$,

$$d\big(g(x_i), g(y_{j,\alpha})\big) \leq \eta. \tag{13}$$

By choosing a specific value of the threshold $\eta$, EER is computed by equaling the false rejection rate (FRR) and the false acceptance rate (FAR). It is straightforward to see that the cases of pitch scaling and VTLN with $\alpha = 0$ degrade to the conventional ASV task without voice disguise.

In order to identify the speaker in disguised voices, a pre-requested step might be to check if the voice is disguised or not. However, in the evaluation recipe presented above, speaker verification is conducted directly without considering if a piece of voice is disguised or not, which is a more difficult task than that only identifying the existence of disguising which is actually a sub-task to check if $\alpha$ is zero. As will be seen later in Section IV, the value of $\alpha$ will be estimated in an automatic way. However, incorporating the proposed method would inevitably result in higher false alarm rates for the underlying ASV system. Therefore, the performance drop of the system will be evaluated and discussed on the data without AVD in Section VI.C. Please refer to the cases with $\alpha = 0$ on Fig. 4.

## IV. RESTORATION OF AUTOMATICALLY DISGUISED VOICES

In order to alleviate the negative impact of voice disguise on ASV, a critical step would be to restore the disguised voice to its original version on which speaker verification is subsequently performed and listening test can also be conducted to give interpretable evidences. As summarized in Section II, AVD by pitch scaling and VTLN have both been summarized by a disguising parameter $\alpha$ in (8). Once an estimation of the disguising parameter is given, say $\hat{\alpha}$, the restored voice is obtained by,

$$\hat{x} = f^{-1}(y; \hat{\alpha}), \tag{14}$$

where $f^{-1}(\cdot)$ is the inverse transform of (3) and (7) for pitch scaling or (9) - (12) for VTLN and $\hat{x}$ is the restored voice. For the case with $i=j$, a relatively low distance is expected for some $\alpha$; while for the case with $i \neq j$, relatively high distances are expected for all $\alpha$'s.

However, for VC, it is difficult to model the disguise by one parameter, therefore its restoration is approximated by taking $f^{-1}(\cdot)$ in forms of pitch scaling or VTLN instead. This is an ad hoc solution to make the restoration of VC disguised voices feasible.

Therefore, the restoration boils down to estimating the parameter $\alpha$. In this section, for pitch scaling, two methods are presented. The first one is the estimation of disguising parameters by using the ratio of fundamental frequencies from [21], which is taken as the baseline in this paper. The other is to use ASV as a tool to estimate the disguising parameters by minimizing a function of ASV scores w.r.t. disguising parameters. To the best of our knowledge, there is no existing solution for the restoration of VTLN, so only our proposed approach is evaluated.

### A. Estimating the disguising parameter by the ratio of fundamental frequencies [21]

In phonetics, pitch is usually used to describe the human's perception of the frequency of sound. AVD conducted by pitch scaling scales the frequency components according to different disguising parameters, thereby also changes the pitch of the voice. The fundamental frequency reflects the important characteristics of the excitation source of speech and is a relatively stable frequency component in the speech signal.

The ratio between the pitch of the original voice, $p_0$, and the pitch of the disguise voice, $p_1$, can be utilized to estimate the disguising parameter as presented in (6) and (7). The method is called *F0-ratio* in short in this paper. The estimated disguising parameter is subsequently applied to restore the MFCC features which are the inputs of ASV to identify the disguised speaker.

For each pair of evaluation, $\{x,y\}$, the following steps are used to make an estimate.

1) A sequence of fundamental frequencies are extracted from the frames of $x$, and their average value $f_x$ is calculated.
2) A sequence of fundamental frequencies are extracted from the frames of $y$ and their average value $f_y$ is calculated. The disguising parameter $\alpha$ is calculated by (15),

$$\hat{\alpha} = 12 \log_2 \left( \frac{f_y}{f_x} \right). \tag{15}$$

In steps 1) and 2), Simplified Inverse Filter Tracking (SITF) is used to extract the fundamental frequency.

3) With the restored spectrum, if necessary, Griffin-Lim algorithm is applied to recover the restored speech [37].
4) Restored voices are then utilized to compute the distance for the trial $\{x,y\}$,

$$d(g(x), g(f^{-1}(y; \hat{\alpha}))). \tag{16}$$

It is worth noting that one does not have to obtain the waveform to compute the distance, but the restored MFCCs should be enough for computing (16).

### B. ASV as a tool to estimate disguising parameters

For each pair of evaluation, $\{x,y\}$, the following steps are used to make a decision, as is shown in Fig. 3.

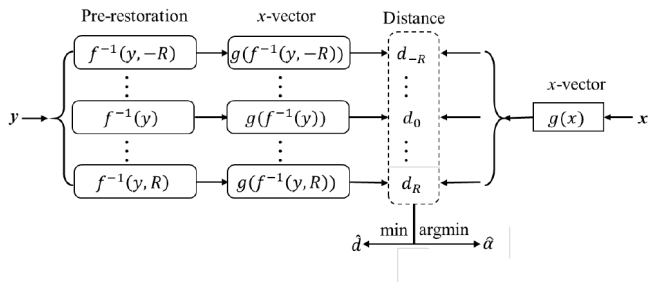

FIG. 3: ESTIMATION OF DISGUISING PARAMETER BY OPTIMIZING ASV. PRE-RESTORATION IS CONDUCTED WITH VARIOUS DISGUISING PARAMETERS FOR A TESTING UTTERANCE. THE PRE-RESTORED VOICES ARE COMPARED WITH THE ENROLLMENT UTTERANCE. THE MINIMAL VALUE AND THE CORRESPONDING PARAMETER ARE COMPUTED AS OUTPUTS.

1) The disguised voice $y$ is firstly pre-restored by a parameter $\alpha$ ranging from $-R$ to $R$, where $R$ is the range of disguising parameter $\alpha$ for pitch scaling or VTLN.

2) Both the re-restored voices $f^{-1}(y;\alpha)$ and the original voice $x$ are fed into TDNN to extract $x$-vectors, $g(f^{-1}(y;\alpha))$ and $g(x)$, respectively.

3) The distances of the pre-restored voices and the original voice are computed by using the recipe $d$ in ASV as explained in Section III.A.

4) Finally, the lowest distance $\hat{d}$, the optimal $\alpha$ denoted by $\hat{\alpha}$, and the restored speech with the optimal $\alpha$, are obtained by solving the following optimization,

$$\hat{d} = \min_{\alpha} d\left(g(x), g\left(f^{-1}(y;\alpha)\right)\right),$$
$$\hat{\alpha} = \arg\min_{\alpha} d\left(g(x), g\left(f^{-1}(y;\alpha)\right)\right). \quad (17)$$

Besides $x$-vectors in Fig.3, any method for feature extraction in ASV is applicable to perform as $g(\cdot)$, e.g. $i$-vector, whose configuration and performance will be reported in Section V and Section VI, respectively.

Given the fact that VC is too complex to be characterized by a couple of parameters, in this paper $f^{-1}(\cdot)$ of VC is approximated by pitch scaling or VTLN instead. This trick can be understood as a low order approximation of the complex transforming functions in VC.

### C. Evaluation of ASV on restored voices with matched and mismatched conditions

With the restored voices and features, a distance for each trial is obtained. By adjusting a threshold, EER is computed to evaluate the restoration of disguised voices in an objective way.

In this paper, evaluations with both matched and mismatched conditions of restorations are considered. The matched condition refers to that the disguising function $f$ and the restoration function $f^{-1}(\cdot)$ take the same format. While for the mismatched one, $f^{-1}(\cdot)$ only takes an approximation, as discussed below.

#### 1) Mismatch within disguising styles

In Section II, both frequency-domain and time-domain disguise are introduced for pitch scaling. However, in the above procedures of the restoration for pitch scaling, no assumption on how the disguised voice $y$ is made. For either kind of disguise, only the reverse transform on spectrum is taken as the means of restoration to evaluate the generalizability of the proposed method. It is worth noting that for time-domain restoration, only the pitch and spectrum can be recovered, and the speed cannot.

For VTLN, only the power function in (11) is taken as the $f^{-1}(\cdot)$ to restore voices disguised by any kind of the four warping functions from (9) to (12) in order to evaluate the mismatch between disguise and restoration.

#### 2) Mismatch between disguising styles

Mutual restoration between pitch scaling and VTLN are also considered to evaluate the generalizability of the proposed approach in Fig.3.

Beside this, in order to make the restoration of VC disguised voices possible, both pitch scaling and VTLN (power) are used to approximate the inverse of the conversion process happened in VC, which is an ad hoc feasible solution to the challenging task.

## V. EXPERIMENTAL SETUP

In this section, detailed experimental setup of datasets, evaluation, baselines are presented.

### A. Datasets with real-world recording conditions

Voxceleb1 [38] and Voxceleb2 [39] are chosen as the primary datasets for evaluating the vulnerability of ASV on AVD and the restoration of AVD by ASV. Both of the datasets are recorded under noisy and reverberant environments, which are close to real-world scenes. Unlike in clean recording conditions, the appearance of miscellaneous noises when conducting AVD bring great challenge to estimate the disguising parameters and to the restoration of original voices. The details of the data for training and testing are listed in Table I.

TABLE I
DETAILED SETUP OF THE TRAINING AND TESTING DATASET

| Task | Dataset | #Speakers | #Utterances |
|---|---|---|---|
| Training | Voxceleb1 dev | 1211 | 21820 |
| | Voxceleb2 dev | 5994 | 145569 |
| | Voxceleb2 test | 118 | 4911 |
| Testing | Voxceleb1 test | 40 | 678 |

The dataset from VCC2018 is used to evaluate our approach on the restoration of VC disguised voices and to analyse the pros and cons of the disguising tools in Section VI.F for a fair comparison. The details of this dataset will be described in Section V.B.

### B. Generation of automatically disguised voice

In order to make the results reproducible, voice disguising of pitch scaling is generated by an open source software *SoundStretch* from [40], which modifies the pitch, rate and tempo of an audio file. Given the fact that the modification of tempo has little impact on speaker verification, this paper only considers changing pitch in frequency-domain disguise and changing rate in time-domain disguise. The disguising parameter $\alpha$ of pitch scaling is ranged from -11 to 11, where only integers are considered. Therefore, the original 678 testing utterances yield 678×23 combinations in total, among which 10k trials are randomly taken for the disguising by pitch scaling.

Per the disguising with VTLN, the open source toolbox in [25] is utilized, again for the reproducibility of our results. The toolbox has four choices of nonlinear warping functions, *bilinear*, *quadratic*, *power* and *piecewise-linear*, to conduct the nonlinear mapping of frequency axes from the original voices to the disguised voices according to different disguising parameters. The range is set as large as possible until the disguised voice is unintelligible, while the step is taken as small as possible unless the disguised voices with neighboring parameters are distinguishable. The values are listed in Table II.

TABLE II
THE RANGE OF THE DISGUISING PARAMETERS FOR VTLN WARPING FUNCTIONS

| Warping function | Range | Step |
|---|---|---|
| **Bilinear** | [-0.3, 0.3] | 0.02 |
| **Quadratic** | [-2, 2] | 0.2 |
| **Power** | [-0.5, 0.5] | 0.05 |
| **Piecewise-linear** | [0.5, 1.5] | 0.05 |

For each trial $\{x,y\}$ to evaluate the disguise conducted by VTLN, $x$ is randomly chosen from the original 678 utterances, while $y$ is randomly chosen from the disguised ones by appointing one of the four warping functions and choosing $\alpha$ randomly in Table II. In this paper, 10k trials are randomly taken for the disguising using VTLN where $x$ and $y$ share the same utterance ID with those in pitch scaling for comparison purpose.

A competitive baseline provided by VCC 2018, *sprocket*, is taken as the tool for VC disguise [28]. *Sprocket* is open source, has well-established routines and has generated ready-for-use samples available online. In this paper, the samples from the target speakers are unused since it is only concerned to recover the identity of the source speaker given his/her converted samples.

There are 280 samples from eight source speakers and 11200 samples converted from the eight speakers, which are treated as original and disguised voices, respectively. For each trial $\{x,y\}$, $x$ and $y$ are randomly chosen from the original and disguised voices, respectively. In this paper, 1.6k trials are randomly taken for evaluation.

### C. Restoration and evaluation setup

As presented in Section IV.C, both pitch scaling and VLTN *power* are utilized to restore the disguised voices generated in Section V.B. For the restoration by pitch scaling, each disguised voice $y$ is pre-restored into 23 versions by enumerating $\alpha$ from -11 to 11. While for the restoration by the power function of VTLN in (11), each disguised voice $y$ is restored into 21 versions by enumerating $\alpha$ from -0.5 to 0.5 with step 0.05.

EERs computed on disguised voices without any restoration are also given to illustrate the threat of AVD on ASV. For the ASV baseline without disguise, the standard list of trials given in Voxceleb1 is used, which has 37,720 trials in total [38].

### D. Time-Frequency features utilized in ASV

After removing the silent parts of an utterance by voice activity detection, a window with 25ms frame-length and 15ms frame-shift is applied to extract acoustic features. Each frame is represented by its MFCC, a 24-dimensional vector with its 1st and 2nd order differences across time, i.e. 72-dimension in total.

### E. i-vector for ASV

The GMM-UBM *i*-vector model used in this paper is trained on the *Training* part of Table I. The number of Gaussian components is 2048. Maximum a posteriori (MAP) is used to adjust the MFCC features of each enrollment/testing utterance to obtain a 400-dimensional vector. Then LDA is utilized to reduce the dimension of the vector from 400 to 200. The 200-dim vector is called *i*-vector. A PLDA is learned as the backend classifier to compute the distance of the two *i*-vectors extracted for each trial $\{x,y\}$.

### F. x-vector for ASV

The training data is again the *Training* part of Table I. The architecture of TDNN is shown in Table III where $T$ is the number of frames of an utterance. The margin in AMSoftmax loss is 0.20. The model is trained from scratch. The batch size is 64 and the learning rate starts with 0.01 and is divided by 10 after 12k iterations without performance promotion. The weight decay parameter is 1.0e-3. The training has finished at around 40k iterations. No data augmentation is applied during training.

Like in the extraction of *i*-vector, LDA is also applied to reduce the dimension of the output of *Segment-Layer* 6 in Table III (i.e. $l_6$ in Fig.2), from a 512-dim vector to a 200-dim one which is called *x*-vector. A PLDA is again learned as the backend classifier to output the distance of the two *x*-vectors extracted for each trial $\{x,y\}$.

TABLE III
THE ARCHITECTURE OF TDNN

| Layer | Layer Context | Structure |
|---|---|---|
| Frame-Layer 1 | $[t-2,t+2]$ | 216×512 |
| Frame-Layer 2 | $\{t-2,t+2\}$ | 1536×512 |
| Frame-Layer 3 | $\{t-3,t+3\}$ | 1536×512 |
| Frame-Layer 4 | $\{t\}$ | 512×512 |
| Frame-Layer 5 | $\{t\}$ | 512×1500 |
| Statistic Pooling | $[0,T)$ | $1500T \times 3000$ |
| Segment-Layer 6 | $\{0\}$ | 3000×512 |
| Segment-Layer 7 | $\{0\}$ | 512×512 |

## VI. RESULTS AND DISCUSSION

In this section, we first report the results of the impact of AVD (i.e. pitch scaling, VTLN and VC) on ASV and the performance of its restoration by ASV. Extensive comparison with an existing solution using F0-ratio is subsequently presented to evaluate the usefulness of the proposed approach. Intuitive illustration of the role of the method on pitch scaling is also given by visualizing the distribution of ASV features. Results on match and mismatch conditions between disguising and restoration are discussed to explore possible improvements on difficult cases e.g. VTLN and VC.

### A. Impact of AVD on ASV and its restoration by ASV

The EERs of the disguise-restoration pairs are given in Table IV for *i*-vector and in Table V for *x*-vector. Some interesting conclusions are drawn from the tables.

TABLE IV
EERs (%) OF ASV ON VOXCELEB1-TEST (I-VECTOR)

| Disguise / Restoration | Pitch scaling | VTLN (all) | Voice conversion |
|---|---|---|---|
| None | 44.14 | 39.86 | 46.80 |
| Pitch scaling | **13.58** | 28.92 | **42.66** |
| VTLN (power) | 39.12 | **24.98** | 43.54 |

TABLE V
EERs (%) OF ASV ON VOXCELEB1-TEST (X-VECTOR)

| Disguise / Restoration | Pitch scaling | VTLN (all) | Voice conversion |
|---|---|---|---|
| None | 38.84 | 34.40 | 43.91 |
| Pitch scaling | **7.10** | 21.54 | **43.54** |
| VTLN (power) | 33.40 | **18.54** | 43.66 |

#### 1) Higher nonlinearity in disguise, greater threat to ASV

By comparing the columns in both Table IV and V, increasing trends of EERs are observed with the increasing of nonlinearity in disguising, i.e. from pitch scaling to VC. Even with voice restoration, disguising methods with higher nonlinearity still yield higher EERs, which indicat greater threat to ASV systems.

#### 2) More advanced ASV, fewer chances deceived by AVD

By comparing the values of the corresponding entries of Table IV and V, *x*-vector generally yields lower EERs than *i*-vector, except the slight fluctuations on VC. This indicates that more advanced ASV approaches, fewer chances deceived by AVD.

#### 3) Matched disguise-restoration performed good

It is not strange that best results are always obtained when the disguise method matched the restoration one, as notified by the bold texts in Table IV and V. However, it is worth noting that even in the matched case the disguising parameter is not known, which is slightly different from the white-box case commonly used in analyzing the vulnerability of artificial intelligent systems, where white-box means both model and parameters are known [41].

#### 4) Linear, quadratic and beyond

Pitch scaling seems a better way than VTLN (power) for completely blind restoration, by comparing the EERs of pitch scaling restoration (i.e. the second rows of Table IV and V) and the EERs of VTLN (power) restoration (i.e. the second rows of Table IV and V).

The combination of the first order pitch scaling and the second order power would further reduce EERs presented in Table IV and V. More sophisticated polynomial functions with higher orders would also work, which is worth exploring as the future work.

### B. Subjective evaluation of AVD and its restoration by ASV

In this section, subjective evaluation of AVD and its restoration by ASV is conducted, where only 30 trials are randomly chosen from the total 10k trials presented in Section V.B to make it a feasible job for human listening. For each of the nine combinations in Table V, a half of trials are from the same speaker while the reaming ones are from different speakers, in order to generate a balanced testing set. Listeners are asked to decide if the two utterances of each trial are from the same speaker or not [1]. Error rates are then calculated and presented in Table VI.

TABLE VI
ERROR RATES (%) OF HUMAN SPEAKER VERIFICATION ON VOXCELEB1-TEST (RESTORATION BY X-VECTOR)

| Disguise / Restoration | Pitch scaling | VTLN (all) | Voice conversion |
|---|---|---|---|
| None | 50.00 | 46.67 | 46.67 |
| Pitch scaling | **13.33** | 40.00 | 50.00 |
| VTLN (power) | 36.67 | **23.33** | **43.33** |

It can be seen from the first row of Table VI that AVD seriously confuses human speaker verification (HSV) where random guess is observed. With the restoration of disguised voices by ASV using *x*-vector, the accuracy of HSV is improved to a certain amount. The relative improvements of restoration on disguise are consistent w.r.t. those in Table V. However, even with restoration, HSV generally performs worse than ASV, which indicates that the necessity of implementing ASV to assist human's decision.

### C. Comparison with the F0-ratio baseline on pitch scaling

The average results of EERs through 23 disguising parameters of pitch scaling are summarized in Table VII. The EER of ASV increases from 2.057% to 38.84%/38.68% with disguised voices generated by frequency-domain/time-domain transform. All restoration methods have reduced the EERs. Compared to its performance on TIMIT reported in [21], F0-ratio does not perform very well on the noisy dataset Voxceleb1, where only 8% EER reduction is observed by yielding an EER around 30%.

Restoration by both *i*-vector and *x*-vector improves the performance of ASV significantly on disguised voices where

---
[1] Audio samples: https://github.com/jiakangli93/disguise-reverse

*x*-vector outperforms *i*-vector. The results again tell us that more advanced ASV methods would yield more accurate restoration results.

The performance per disguising parameter is shown in Fig.4. Without restoration, larger disguise parameter brings more distortion to original voices, so EER increases rapidly from $\alpha=0$ to $|\alpha|=11$. The restoration by F0-ratio, *i*-vector and *x*-vector yields relatively consistent improvements through all the values of the disguising parameter w.r.t. the ASV without restoration. It is clear to see that at some cases, the performance of *x*-vector approximates its lower bound of *x*-vector without disguising, say $\alpha=-4$ and $\alpha=2$.

TABLE VII
RESULTS OF ASV ON VOXCELEB1-TEST

| | Model | EER (%) |
|---|---|---|
| | x-vector without disguise | 2.057 |
| Time-Domain Disguise | x-vector without restoration | 38.68 |
| | restoration with F0-ratio | 30.50 |
| | restoration with *i*-vector | 14.28 |
| | restoration with *x*-vector | **7.540** |
| Frequency-Domain Disguise | x-vector without restoration | 38.84 |
| | restoration with F0-ratio | 30.96 |
| | restoration with *i*-vector | 13.58 |
| | restoration with *x*-vector | **7.100** |

The result of ASV without restoration when $\alpha=0$ (the red-circled line in Fig.4) is tested on the 10k trials as is presented in Section V.C. However, the baselines of ASV without disguise (i.e. the black lines in Fig.4) are obtained on the default testing set of Voxceleb1 with 37,720 trials. This makes the two results cannot be compared directly. However, we find that when $\alpha=0$ voice restoration did not work based on the observation that the pink-diamond dash-dotted line (i.e. ASV with restoration) is higher than the red-circled one (i.e. ASV without restoration) near $\alpha=0$ in both (a) and (b) of Fig.4. This degradation of performance may come from the inaccurate estimation of disguising parameters when making the restoration, i.e. false alarm has been introduced.

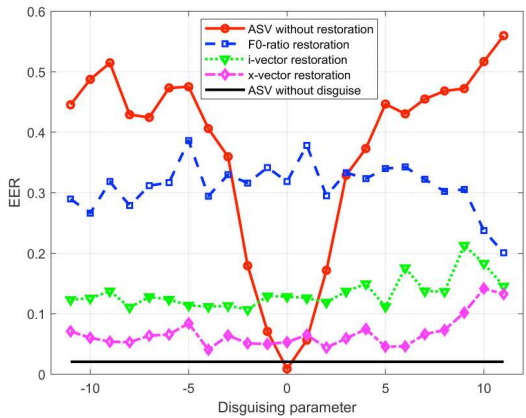

(A) FREQUENCY-DOMAIN DISGUISING

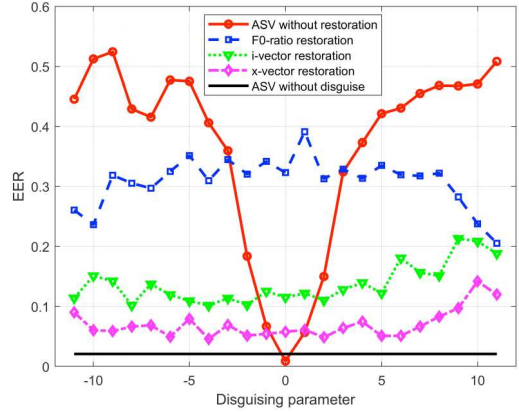

(B) TIME-DOMAIN DISGUISING

FIG. 4. EER OF ASV WITH/WITHOUT RESTORATION OF PITCH SCALING DISGUISE FOR DISGUISING PARAMETERS RANGING FROM -11 TO 11.

For further investigating the performance of voice restoration, the statistical performance on the estimation of disguising parameter is shown in Fig.5, where the error bars are plotted for the bias values of the estimations over all the trials in the list. The ideal result should always be zero. Form the error bars of the bias values, it is clearly seen that *x*-vector yields results with both low bias and low variances.

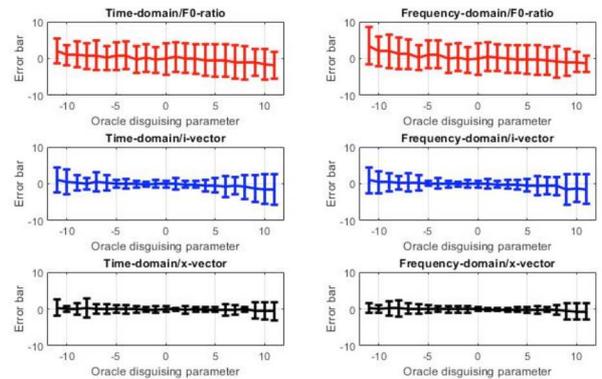

FIG. 5. ERROR BARS OF THE BIAS VALUES OF THE ESTIMATED DISGUISING PARAMETER V.S. THE ORACLE VALUE IN PITCH SCALING. THE IDEA RESULT SHOULD BE ALWAYS ZERO.

Special analysis on spectrogram is conducted to explain the unsymmetrical performance of voice restoration w.r.t. disguising parameters in Fig.4 where worse performance on $\alpha>0$ is observed than that on $\alpha<0$. It is found that $\alpha>0$ implies the stretching of frequency axis as seen from (a) to (b) of Fig. 6, so the restoration is a reverse process by compressing the frequency axis. Therefore, values in some high-frequency bands are missing and should be filled by some values as shown in (c) and (d) of Fig. 11, which bring errors compared to the true values. This explains the phenomenon that EERs on positive disguising parameters are relatively larger than those on negative disguising parameters as observed in Fig.4.

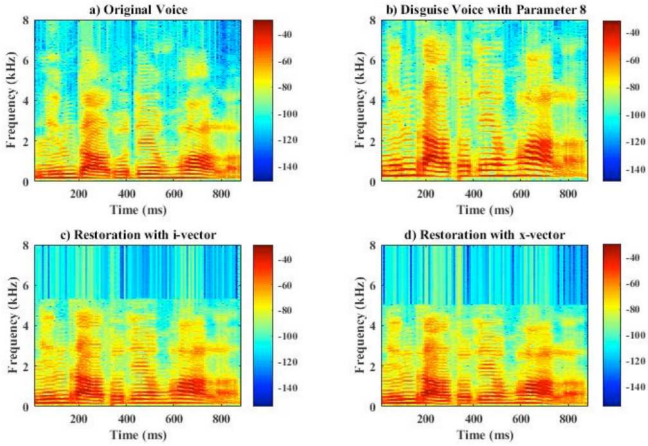

FIG. 6. THE SPECTROGRAM OF AN UTTERANCE. (A) ORIGINAL VOICE, (B) DISGUISED VOICE WITH DISGUISING PARAMETER 8, (C) RESTORED SPECTROGRAM WITH I-VECTOR, AND (D) RESTORED SPECTROGRAM WITH X-VECTOR.

### D. Intuitive illustration of the impact of AVD on ASV

Four speakers each with three utterances are chosen to investigate the role of AVD and its restoration. The $i$-vectors and $x$-vectors of the 12 utterances with the disguising of pitch scaling and VTLN (power) and their restorations are extracted for analysis. t-SNE is taken as the dimension reduction tool to visualize the $i$-vectors (Fig. 7) and $x$-vectors (Fig. 8). From both figures, it is found that, although the original utterances (denoted by dots with four different colors, each color for each speaker) of the four speakers, are separated from each other, their disguised voices (denoted by triangles with four different colors) spread over the whole space. The spreading of the $i$-vectors or $x$-vectors has reduced the between-class distances and has enlarged the within-class distances, which makes ASV extremely difficult to distinguish different speakers. The larger the absolute value of the disguising parameter is, the further $i$-vectors and $x$-vectors drift away from the original ones without disguise. This also explains why the EERs of ASV increases rapidly with the increasing of the absolute value of disguising parameters in Fig.4 for the case of pitch scaling.

By inspecting Fig. 7 and Fig. 8, it is straightforward to see the role of voice restoration by using $i$-vector and $x$-vector. The margins among the four speakers are significantly enlarged as seen from (a) to (b) and (c) to (d) in both Fig. 7 and Fig. 8. In fact, voice restoration by either $i$-vector or $x$-vector condenses the features of each speaker by pulling the features of disguised voices back to the ones of its original version.

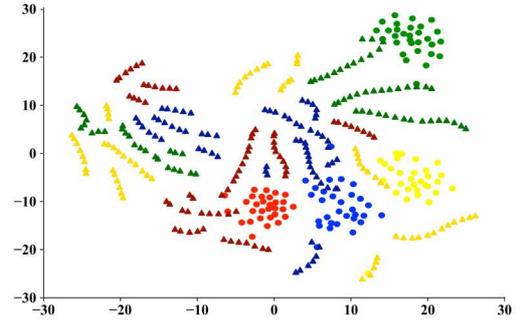

(A) DISGUISED BY PITCH SCALING

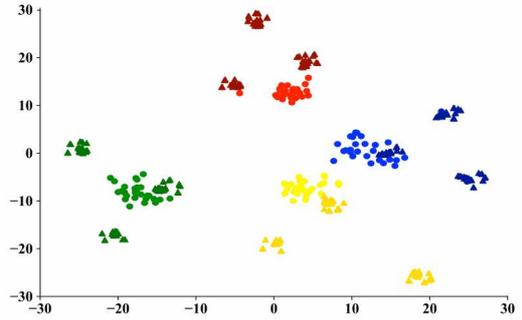

(B) RESTORED BY PITCH SCALING

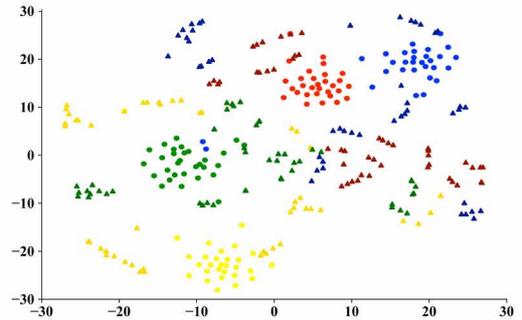

(C) DISGUISED BY VTLN (ALL)

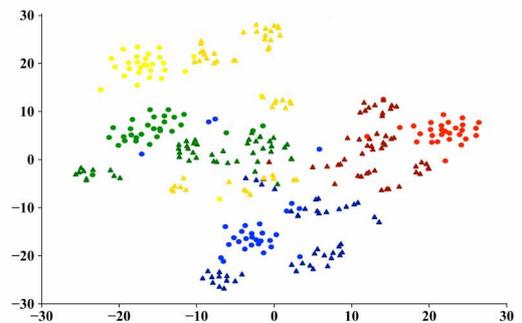

(D) RESTORED BY VTLN (POWER)

FIG. 7. A T-SNE FIGURE TO VISUALIZE I-VECTORS OF THE ORIGINAL, DISGUISED, AND RESTORED VOICES FROM 4 DIFFERENT SPEAKERS. THE DOTS

WITH FOUR LIGHT COLORS (I.E. RED, YELLOW, GREEN, AND BLUE) REPRESENT THE FEATURES OF THE ORIGINAL VOICES OF THE FOUR SPEAKERS RESPECTIVELY. THE TRIANGLES WITH FOUR DARK COLORS (I.E. DARK RED, DARK YELLOW, DARK GREEN, AND DARK BLUE) REPRESENT THE FEATURES OF THE DISGUISED VOICES OF THE FOUR SPEAKERS. (BETTER VIEWED IN COLOR)

With a careful comparing of Fig. 7 and Fig. 8, it is clearly seen that $x$-vector does a better job than $i$-vector by yielding a more compact cluster of ASV features for each speaker, which justifies the good performance of $x$-vector w.r.t. $i$-vector in Table IV, V, VII and Fig.4.

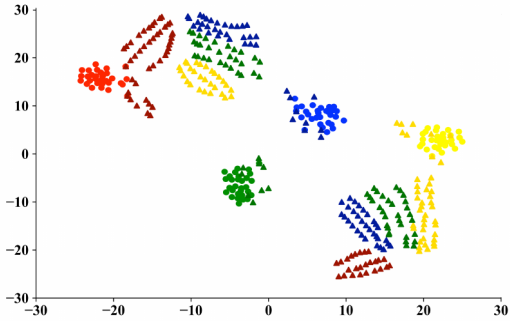

(A) DISGUISED BY PITCH SCALING

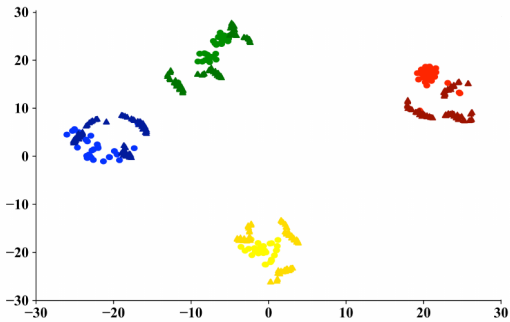

(B) RESTORED BY PITCH SCALING

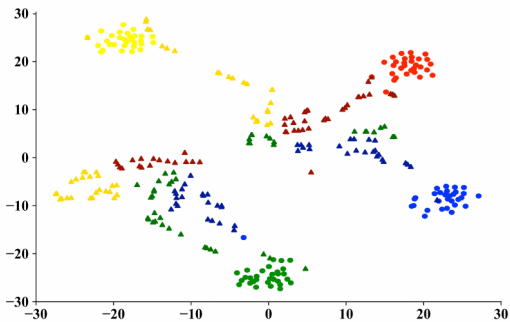

(C) DISGUISED BY VTLN (ALL)

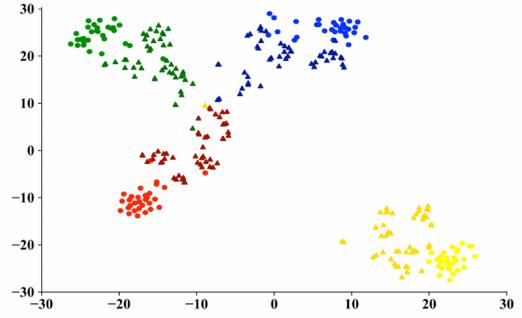

(D) RESTORED BY VTLN (POWER)

FIG. 8. A T-SNE FIGURE TO VISUALIZE **X-VECTORS** OF THE ORIGINAL, DISGUISED, AND RESTORED VOICES FROM 4 DIFFERENT SPEAKERS. THE DOTS WITH FOUR LIGHT COLORS (I.E. RED, YELLOW, GREEN, AND BLUE) REPRESENT THE FEATURES OF THE ORIGINAL VOICES OF THE FOUR SPEAKERS RESPECTIVELY. THE TRIANGLES WITH FOUR DARK COLORS (I.E. DARK RED, DARK YELLOW, DARK GREEN, AND DARK BLUE) REPRESENT THE FEATURES OF THE DISGUISED VOICES OF THE FOUR SPEAKERS. (BETTER VIEWED IN COLOR)

### E. Approximation of the nonlinear warping in VTLN

The detailed results of the restoration of all the four kinds of VTLN mapping functions in (9) -(12) by both pitch scaling and VTLN (power) are given in Table VIII. It is seen from the table that power function approximates well to *bilinear*, *quadratic* and *power* itself, given their similarity on warping curves. Pitch scaling works well for *piecewise-linear*, which is not surprising given their linearity in common. The findings will further be explained in Fig.9 and 10.

TABLE VIII
EERs (%) OF VTLN DISGUISED VOICES RESTORED BY PITCH SCALING / POWER FUNCTION

| Disguise function | $i$-vector | $x$-vector |
|---|---|---|
| Bilinear | 26.80 / **24.94** | 18.97 / **17.11** |
| Quadratic | 27.80 / **19.66** | 19.05 / **12.75** |
| Power | 38.39 / **19.48** | 33.93 / **13.96** |
| Piecewise-linear | **21.03** / 35.70 | **12.96** / 28.61 |

By taking some examples to inspect the disguising function and its corresponding restored function estimated by the proposed approach in Fig.3, it is seen that power function is a good approximation to bilinear, quadratic, and power itself, as seen from (a)-(c) of Fig. 9. However, there is a gap between the piecewise linear function and its approximation as seen from Fig.9 (d).

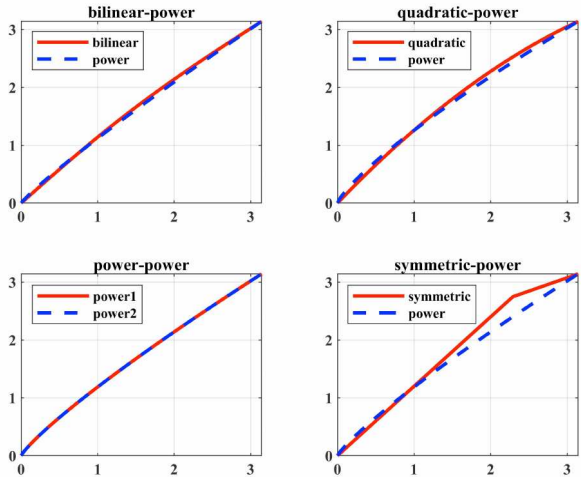

FIG. 9: COMPARISON OF THE FOUR KINDS OF WARPING FUNCTIONS IN VTLN DISGUISING (RED LINES) AND THEIR CORRESPONDING RESORTED APPROXIMATIONS BY THE **POWER** FUNCTION (BLUE DASHED LINES), WHERE 'A-B' REFERS TO AVD BY FUNCTION A AND RESTORATION BY FUNCTION B.

The approximation of the four VTLN functions by pitch scaling is shown in Fig.10 where the pitch scaling parameter is again estimated by the approach in Fig.3. The estimated lines perform as tangents of the four functions with emphasis on the low frequency bands.

Given the intuitive interpretation above, it seems reasonable to deploy low order functions (e.g. the first-order pitch scaling or the second-order power function) to approximate nonlinear mappings in AVD. It is interesting to study other forms of approximation functions as well as retaining the solvability of (17).

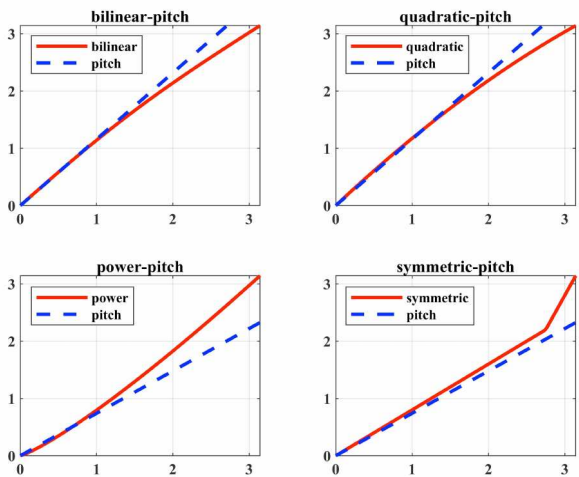

FIG. 10: COMPARISON OF THE FOUR KINDS OF WARPING FUNCTIONS IN VTLN DISGUISING (RED LINES) AND THEIR CORRESPONDING RESORTED APPROXIMATIONS BY **PITCH SCALING** (BLUE DASHED LINES), WHERE 'A-B' REFERS TO AVD BY FUNCTION A AND RESTORATION BY FUNCTION B.

### F. Pros and cons of the three disguising methods

In the analysis above, VC seems a good choice of AVD to deceive ASV. However, in real-world usage, the performance on deceiving ASV systems is only one factor to be considered. Besides this, both additional computational cost to perform the disguise and the robustness on voice quality should also be considered. We thus studied the mean computational cost to conduct each kind of AVD on the same computer and the mean opinion scores (MOS) to evaluate the quality of the disguised voices.

Without bothering to retrain a VC model, the experiments are carried out on the dataset of VCC2018, where 35 sentences from the same speaker with a total length of 111.39 seconds are chosen to conduct AVD. In order to evaluate the noise robustness of the three AVD methods, white noises are added to the original clean voices with a signal-noise ratio 20dB. The computing platform is a workstation with two Intel Xeon E5V3 CPUs with 2.6G frequency. MOS scores are calculated by 5 listeners between 1 and 5. Real Time Factor defined by the ratio of CPU time over voice duration is utilized to evaluate the computational cost of the disguising methods [42]. When RTF < 1, the system is able to process speech in real time on the computer. The scores on RTF and MOS are presented in Table IX.

TABLE IX
THE COMPLEXITY AND VOICE QUALITY OF THE DISGUISING TOOLS

| Disguise tool | Pitch scaling | VTLN | VC |
|---|---|---|---|
| **RTF** | 0.02 | 0.22 | 1.10 |
| **MOS (1~5)** | 3.90 | 3.00 | 4.10 |

Given the experiments above on the three AVD methods, pitch scaling makes a good compromise on the three evaluation measures. However, with the continuous improvements on VC and the acceleration of scientific computing, we believe VC would be a good choice to AVD and will pose great threat to public security.

In order to improve the defense against the AVD conducted by VC, high level paralanguage characteristics of the disguised speaker from forensic speaker recognition in [43] and the combination of automatic and acoustic–phonetic approaches in [44] may also be helpful. Features regarding duration, intensity, vowel formant frequencies, and long-term average spectrum have been successfully applied to whispery voice disguise in Mandarin Chinese [45]. Trivial events, such as cough, laugh and sniff, have also been applied to improve the forensic speaker examination of human disguise [7]. All these works on forensic speaker recognition with human disguise shed a light on the defense against the AVD by VC.

### VII. CONCLUSION

In this paper, the mutual relation of AVD and ASV was extensively studied and the conclusion was drawn that ASV was not only a victim of AVD but could also be a tool to beat some simple types of AVD.

AVD methods including pitch scaling, VTLN and VC were firstly presented and analyzed. State-of-the-art ASV was then introduced to evaluate the impact of AVD on ASV, where significant performance drop was seen for ASV on disguised voices with all the three disguising types. A method for restoring disguised voice to its original version was subsequently proposed by optimizing ASV scores w.r.t. the parameters of restoration functions, which significantly outperformed related baselines on the disguising with pitch scaling. The proposed approach generalized well on some mismatched conditions, e.g. restoring the pitch scaling AVD by VTLN (power), and vice versa. Visualization of ASV features found that the proposed approach actually pulled the disguised ASV features from the same speaker closer while pushed the features from different speakers further.

However, the current approach failed to cope with complex mapping functions in VC. In future, we would like to explore high order restoration functions to improve the proposed approach's performance on AVD with more sophisticated transforms. Analysis techniques from forensic speaker recognition should also be useful to improve the defense against the AVD conducted by VC.